\documentclass[%
 reprint,
superscriptaddress,
 amsmath,amssymb,
 aps,
]{revtex4-2}

\usepackage{graphicx}% Include figure files
\usepackage{dcolumn}% Align table columns on decimal point
\usepackage{bm}% bold math
\usepackage{hyperref}% add hypertext capabilities
\usepackage{physics}

\begin{document}

\title{Optimal Control of Spin Squeezing in 2D Finite-Range Interacting Systems}

\author{Ang Li}
\affiliation{State Key Laboratory of Low Dimensional Quantum Physics, Department of Physics, Tsinghua University, Beijing 100084, China}

\author{Ling-Na Wu}
\email[]{lingna.wu@hainanu.edu.cn}
\affiliation{Center for Theoretical Physics \& School of Physics and Optoelectronic Engineering, Hainan University, Haikou 570228, China}

\author{Li You}
\email[]{lyou@mail.tsinghua.edu.cn}
\affiliation{State Key Laboratory of Low Dimensional Quantum Physics, Department of Physics, Tsinghua University, Beijing 100084, China}

\date{\today}% It is always \today, today,

\begin{abstract}
Spin squeezing serves as both a fundamental witness of quantum entanglement and a critical resource for quantum-enhanced metrology. While generating substantial spin squeezing in finite-range interacting systems remains challenging, such capability is important for advancing quantum technologies. 
In this work, we develop an optimal control strategy for achieving enhanced spin squeezing in a two-dimensional XX model with dipolar interactions. 
Leveraging rotor-spin-wave theory for periodic boundary conditions, we circumvent computational bottlenecks to explore control strategies at unprecedented scales. Remarkably, optimizing a single collective transverse field is sufficient to achieve substantial squeezing enhancement, exceeding the two-axis-twisting benchmark.
% By engineering a time-dependent transverse field, we achieve squeezing that outperforms the two-axis-twisting benchmark. 
The optimized control field achieves this breakthrough by dynamically suppressing inter-subspace mixing induced by the finite-range interactions, thereby confining the system evolution predominantly within the maximal spin subspace. 
We further extend rotor–spin-wave theory to open boundary conditions and incorporate dephasing noise, providing a scalable framework for realistic systems. Under these conditions,
the optimized protocol remains effective, highlighting its robustness and suitability for experimental implementation.
% Crucially, 
% the scheme remains robust under realistic noise conditions. 
% To overcome the computational intractability of open-boundary system optimization, we develop an efficient approach based on the collective spin model that delivers near-optimal control fields with dramatically reduced computational overhead.
\end{abstract}

\maketitle

%\tableofcontents

%\section{\label{sec:intro} Introduction} 

\textit{Introduction}---
Spin squeezing~\cite{kitagawa1993squeezed}, as a unique and important form of quantum entanglement, reduces quantum fluctuations in a specific spin component below the standard quantum limit by redistributing quantum noise. It plays a crucial role in the characterization of quantum entanglement and quantum precision measurements~\cite{ma2011quantum,Luca18rmp}, holding significant theoretical and practical value.  
Over the past few decades, spin-squeezed states~(SSSs) have been successfully prepared in various experimental platforms, including atomic ensembles in optical cavities~\cite{greve2022entanglement}, ion chains~\cite{Bohnet16quantum}, and Bose-Einstein condensates~\cite{Gross10nonlinear,Riedel2010atom}. 
% Traditionally, the generation of spin-squeezed states relies on all-to-all interactions (where every particle interacts with the same strength). 
% However, there are many systems that possess finite-range interactions~(where interaction strength decays with distance).

The conventional framework for generating SSSs requires all-to-all interactions, characterized by uniform coupling strengths between all particle pairs. This paradigm enables efficient squeezing protocols such as one-axis twisting (OAT)~\cite{kitagawa1993squeezed}, which rely critically on the system's permutation symmetry. However, many emerging quantum platforms such as Rydberg atom arrays~\cite{browaeys2020many,wu2021concise} and dipolar quantum gases~\cite{DeMille2024,Chomaz2022Dipolar} exhibit finite-range interactions with pronounced spatial dependence, where the coupling strength $J_{ij}$ decays as a function of inter-particle distance $r_{ij}$ (typically following power-law $1/r_{ij}^\alpha$). This breakdown of global symmetry introduces complex spatial correlations that degrade squeezing performance and necessitates new approaches to maintain metrological advantage.
Consequently, developing efficient methods for SSS generation in finite-range interacting systems represents both a fundamental challenge in quantum control and a practical requirement for emerging quantum technologies.

Research in this field has focused on achieving \emph{scalable squeezing}, where the quality of squeezing improves with increasing system size. 
Early studies on power-law interacting Ising models~\cite{Foss2016arXiv} established that scalable squeezing requires long-range interactions ($\alpha < D$, where $D$ is the spatial dimension). Subsequent studies of $U(1)$-symmetric systems (e.g., XX~\cite{Comparin2022PRA,Comparin2022PRL} and XXZ~\cite{Rey2020PRL,Block2024NP,roscilde2024scalable} models) revealed the  possibility of scalable squeezing even with \emph{short-range} interactions, enabled by either unique low-energy eigenstates structures~\cite{Comparin2022PRA} or finite-temperature symmetry breaking~\cite{Block2024NP,roscilde2024scalable}. However, the squeezing of these models are ultimately constrained by the OAT limit. 
Although two-axis twisting (TAT) interactions in all-to-all interacting systems can surpass this limit and generate stronger squeezing faster, this enhancement crucially depends on global interactions---a direct generalization to finite-range systems fails to break the OAT barrier~\cite{Block2024NP}.

Recent studies utilizing quantum variational circuits have demonstrated the potential for generating high-quality SSSs through optimized sequences of transverse fields and interaction terms~\cite{Rey2019PRL,Fan2023PRR,Castro2024PRA,Carrera2025}. 
However, theoretical studies of such optimization protocols in finite-range interacting systems face severe computational constraints, restricting analyses to two limiting cases: (i) small systems amenable to exact diagonalization~\cite{Fan2023PRR,Castro2024PRA,Carrera2025}, or (ii) one-dimensional systems with rapidly decaying interactions~\cite{Rey2019PRL,Carrera2025}, i.e.~nearest-neighbor couplings or power-law interactions with $\alpha =6$, where matrix product state (MPS) methods remain tractable. Consequently, there exists a significant knowledge gap regarding optimal control strategies for higher-dimensional systems with longer-range interactions, which are more favorable in generating SSSs~\cite{Rey2020PRL}.

In this work, we present a scalable optimal-control strategy for generating strong spin squeezing in a 2D XX model with power-law interactions ($\alpha=3$), exploiting rotor–spin-wave (RSW) theory~\cite{roscilde2023rotor} to overcome the computational challenges of optimizing large systems under periodic boundary conditions. Remarkably, by optimizing only a single collective transverse field, we achieve squeezing that surpasses the two-axis-twisting (TAT) benchmark. The optimized control field achieves this breakthrough by dynamically suppressing inter-subspace mixing induced by the finite-range interactions, thereby confining the system evolution predominantly within the maximal spin subspace. 
We further examine the robustness of the protocol against dephasing noise. The protocol is found to remain advantageous over the uncontrolled case even under relatively strong dephasing.
% , revealing a trade-off between coherent squeezing generation and decoherence: while longer evolution improves squeezing, it also increases susceptibility to noise, resulting in an optimal intermediate timescale where squeezing is maximized.
Finally, to explore the applicability of our approach to more realistic experimental settings, we extend the RSW framework to open boundary conditions and demonstrate that  enhanced squeezing persists in large open systems, highlighting both the robustness and experimental relevance of our approach.

\textit{Model and methods}---We consider a two-dimensional XX spin model with power-law interactions and a temporally varying uniform transverse field, described by the Hamiltonian~($\hbar=1$ hereafter):
\begin{equation}
    H = -\sum_{i<j}J_{ij}(S_i^xS_j^x+S_i^yS_j^y) - h(t) \sum_i S_i^x,
    \label{eq:XX_Ham}
\end{equation}
where the interaction strength decays as a power law $J_{ij} = 4J|\bm{r}_i - \bm{r}_j|^{-\alpha}$, and $S_i^{x,y,z}$ denotes the spin operators for the $i$-th site. This model has been experimentally realized in several quantum platforms: trapped ion systems demonstrate the full range of tunability ($0<\alpha<3$)~\cite{franke2023quantum}, while Rydberg atom arrays naturally realize the $\alpha=3$ case~\cite{bornet2023scalable}. 
In the following, we focus on the $\alpha=3$ case~(dipole-dipole interaction). 
%The more experimentally demanding case of spatially inhomogeneous control, requiring individual site addressing techniques, is left for future work. 

We are interested in the squeezing dynamics of the system starting from a coherent spin state with all spins pointing to $x$ direction, $|\theta=\pi/2,\phi=0\rangle \equiv [\cos(\theta/2) |\uparrow\rangle +\sin(\theta/2) e^{i\phi}|\downarrow\rangle ]^{\otimes N}$. The squeezing performance of the system is quantified by the spin squeezing parameter~\cite{Wineland94squeezed,Wineland92spin}
\begin{equation}
\xi^2 = \frac{N(\Delta S_{\perp,\text{min}})^2}{|\langle \bm{S}\rangle|^2},
\label{eq:SSP_definition}
\end{equation}
where $N$ is the number of spins, $\bm{S} \equiv (S_x, S_y, S_z)$ with $S_{\mu}=\sum_i S_i^\mu$ denoting the collective spin operator and $(\Delta S_{\perp,\text{min}})^2$ represents the minimal variance perpendicular to the mean spin direction. 
The initial coherent spin state satisfies $\xi^2=1$, representing the standard quantum limit. The emergence of spin squeezing ($\xi^2<1$) signals the formation of many-body entanglement that enables sub-shot-noise metrological precision. 
%This parameter elegantly interpolates between the shot-noise limit ($\xi^2=1$) and the Heisenberg limit ($\xi^2 \sim 1/N$), with values $\xi^2<1$ indicating entanglement-enabled metrological advantage.
For numerical optimization, we parameterize the control field $h(t)$ as a piecewise constant function and perform the minimization of $\xi^2$ via gradient-based optimization (\texttt{SciPy}'s \texttt{BFGS} routine~\cite{2020SciPy-NMeth}). 

\begin{figure}[h]
{\includegraphics[width=\linewidth]{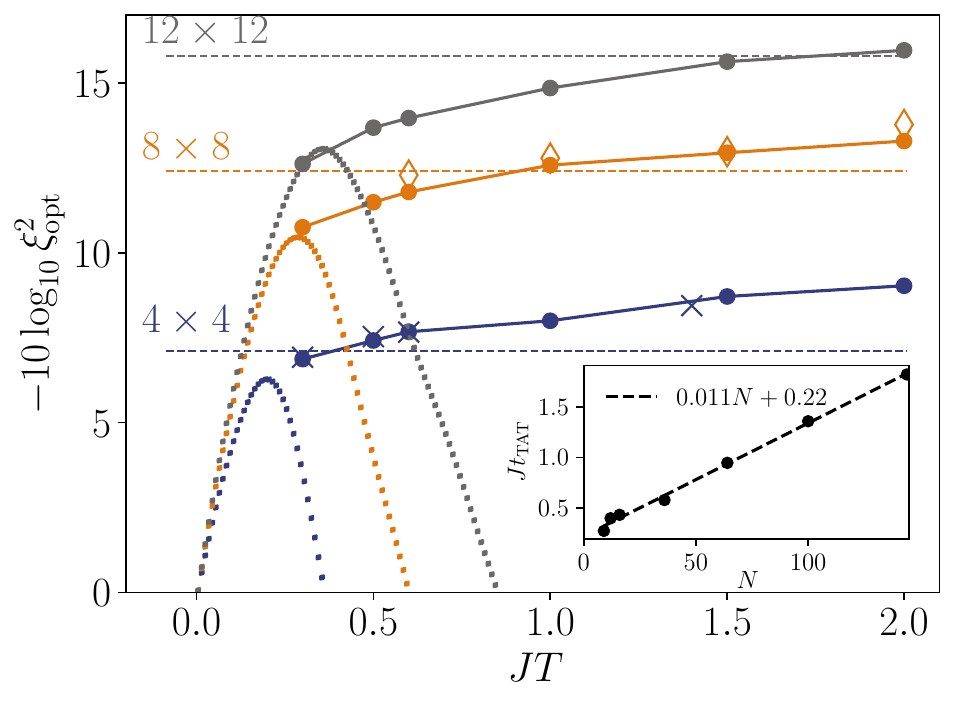}}%
\caption{
The optimal squeezing parameter for the control protocol as a function of total evolution time $T$ for different system sizes under periodic boundary conditions (PBC). Filled circles denote results obtained using the RSW-based optimization, while cross markers (‘x’) and diamonds correspond to exact diagonalization (ED) and time-dependent variational Monte Carlo (t-VMC) results, respectively.
% The optimized squeezing parameter for the control protocol, calculated using the RSW theory (filled circles), ED (cross markers `x'), and time-dependent Variational Monte Carlo (t-VMC, diamonds), is plotted as a function of total evolution time \( T \) for different system sizes with periodic boundary conditions (PBC). 
For comparison, the time evolution of the squeezing parameter for the uncontrolled case (i.e., \( h(t) = 0 \), obtained by RSW theory) is shown as dotted lines, while the TAT limit is indicated by dashed lines. The solid lines connecting the filled circles serve as visual guides.
The inset displays the crossover time \( t_{\mathrm{TAT}} \)---the point at which the optimized squeezing parameters exceed those of the TAT models---as a function of system size.}
\label{fig:rotor_opt_vs_size}
\end{figure}

% \begin{figure}[h]
% {\includegraphics[width=\linewidth]{figure/rotor_opt_vs_size_and_intersection_time_with_TAT.pdf}}%
% \caption{\label{fig:epsart} The optimized squeezing parameter calculated by the RSW theory or ED for different system sizes with PBC. (a). The maximal value of the optimized SSP $-10\log_{10}\Tilde{\xi}^2(t)$ during the time duration $t\in[0,T]$, with $(L_x,L_y)=(3,3),(4,3),(4,4),(6,6),(8,8),(10,10),(12,12)$ for lines or markers from below to above. The dashed lines represent the time evolution $-10\log_{10}\xi^2(t)$ of the uncontrolled cases, and the dot-dashed lines represent the maximum of the SSP for the TAT models with the same system size. (b)The times $t_\mathrm{TAT}$ where the optimized SSPs surpass those of the TAT models.}
% \label{fig:rotor_opt_vs_size}
% \end{figure}

Accurate numerical simulation of 2D lattice spin systems with long range interactions faces a fundamental computational bottleneck—the exponential scaling of Hilbert space dimension with system size becomes prohibitive for optimization tasks demanding both precision and efficiency. For the XX model with periodic boundary conditions, the RSW theory~\cite{roscilde2023rotor} provides an elegant resolution by decoupling the dynamics into two computationally tractable sectors: (i) a collective zero-momentum sector described by an effective OAT rotor [with Hamiltonian $\mathcal{P}_{J=N/2} H \mathcal{P}_{J=N/2}=\chi S_z^2+\mathrm{const}$, effective interaction strength $\chi=2\sum_{i<j} J_{ij}/(N-1)/N$] in a drastically reduced ($N+1$)-dimensional Dicke subspace, and (ii) finite-momentum excitations captured by linear spin-wave theory through O($N$) bosonic correlators~(see Sec.~\ref{app:RSW} of the Supplemental Material for more details). This decomposition, rigorously valid when finite-momentum excitations remain perturbative, enables efficient calculation of squeezing dynamics.

%\section{RESULTS}
% \subsection{Periodic Boundary Condition}
{\textit {Optimal squeezing under periodic boundary condition}}---
Figure~\ref{fig:rotor_opt_vs_size} showcases the central result of our control protocol for systems under periodic boundary conditions: the optimal squeezing parameter (in dB), $-10\log_{10}{\xi}^2_{\rm opt}$, as a function of the total evolution time $T$ for various system sizes.
The reference uncontrolled cases [$h(t)\equiv0$] are indicated by dotted lines, highlighting the quantum control advantage.
% , with dashed lines indicating the uncontrolled~(uncontrolled) case, i.e. $h(t)=0$. 
The RSW theory's validity is evidenced by its excellent agreement with the results from exact diagonalization (ED, indicated by cross markers `x', obtained by Python package \texttt{QuSpin}~\cite{10.21468/SciPostPhys.2.1.003, 10.21468/SciPostPhys.7.2.020}) for small systems and time-dependent variational Monte Carlo~(t-VMC, diamonds) simulations for larger ones~(see Sec.~\ref{app:verification} of the Supplemental Material). 
In the uncontrolled case (dotted lines), squeezing saturates at intermediate times, while optimal control immediately boosts squeezing when the total evolution time $T$ exceeds this saturation point. Most remarkably, the controlled system consistently surpasses the TAT limit (dashed lines) imposed by Hamiltonian $H_{\rm TAT}=\chi(S_z^2 - S_y^2)$---a breakthrough achieved solely through tuning a uniform transverse field. 
The crossover time $t_{\rm TAT}$ increases only weakly with system size, following a favorable linear scaling: $Jt_{\rm TAT} = 0.01N + 0.22$, as shown in the inset.
% , with the crossover time $t_\mathrm{TAT}$ slowy increases with system size, following a linear scaling with the system size: $t_\mathrm{TAT} = 0.011N + 0.22$, as shown in the inset.

\begin{figure}[h]
{
\includegraphics[width=\linewidth]{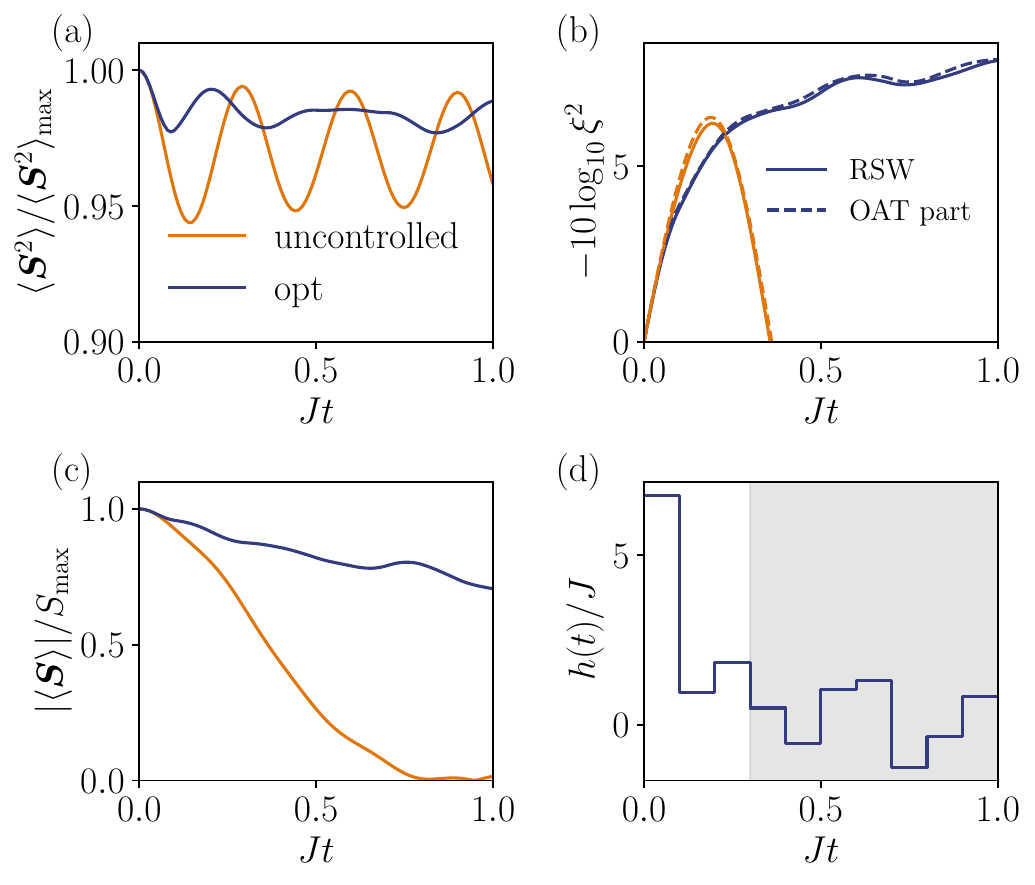}\\
\includegraphics[width=\linewidth]{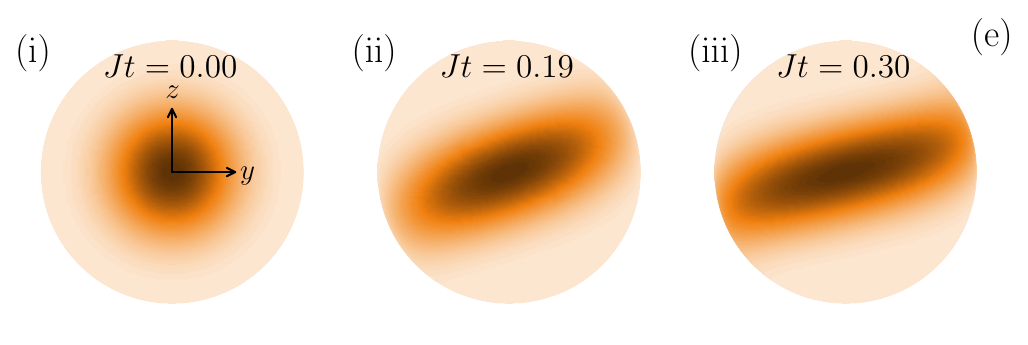}\\
\vspace{-8pt}
\includegraphics[width=\linewidth]{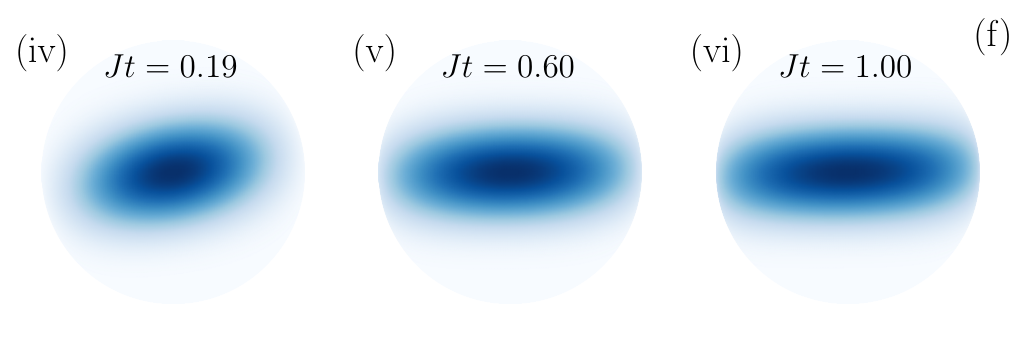}
\vspace{-16pt}
}
\caption{ Time evolution of spin system dynamics under optimal control. 
    (a) Normalized spin squared expectation value $\langle\bm{S}^2\rangle/\langle\bm{S}^2\rangle_{\mathrm{max}}$ showing the optimized (blue) versus uncontrolled evolution (orange). 
    (b) Spin squeezing parameter $-10\log_{10}(\xi^2)$ (in dB), with solid lines representing exact evolution and dashed lines showing effective collective spin models. 
    (c) Normalized mean spin $|\langle\bm{S}\rangle|/S_{\mathrm{max}}$. 
    (d) Optimized step-wise constant control field $h(t)$. All quantities are plotted versus dimensionless time $Jt$, where $J$ represents the interaction strength. 
    (e)-(f)~Visualization of the spin squeezing dynamics in (b) on the Bloch spheres for the uncontrolled case and optimized case, respectively. We look down at the $y-z$ plane from the positive $x$-axis.}
\label{fig:Bloch_rotor_Lx_8_Ly_8}
\end{figure}

To unravel the quantum control mechanism, we monitor the spin dynamics through the time evolution of the total spin squared operator $\langle {\bm {S}}^2\rangle \equiv \langle S_x^2 + S_y^2 + S_z^2\rangle$, as shown in Fig.~\ref{fig:Bloch_rotor_Lx_8_Ly_8}(a) for a $4 \times 4$ system with $T = 1.0/J$. In the case of all-to-all interactions, $\langle {\bm S}^2 \rangle$ is conserved at $S_{\rm max}(S_{\rm max}+1) = N(N/2 + 1)/2$, reflecting confinement to the maximal spin sector with the total spin $S_{\rm max} = N/2$. For finite-range interactions, however, deviations from perfect SU(2) symmetry permit leakage into lower spin sectors and induce the consequent decay of $\langle {\bm {S}}^2 \rangle$, which fundamentally limits achievable spin squeezing.
Figure~\ref{fig:Bloch_rotor_Lx_8_Ly_8}(a) demonstrates this effect clearly: The uncontrolled system (orange data) exhibits oscillatory decay of $\langle {\bm S}^2\rangle$, signaling the inter-subspace coupling.
In comparison, the controlled system~(blue data) shows suppressed decay of $\langle {\bm S}^2\rangle$, demonstrating the field's
active suppression of detrimental subspace transitions.
This is further evidenced by the steady growth of spin squeezing under control~[Fig.~\ref{fig:Bloch_rotor_Lx_8_Ly_8}(b), blue curve] and the remarkable agreement between exact numerical results~(solid lines) and predictions from the effective collective spin model that neglects inter-subspace coupling~(dashed lines).

%full RSW theory~(incorporating both collective spin dynamics and excitations) and the pure collective spin model, confirming the suppression of inter-subspace coupling. 

% However, finite-range interactions introduce leakage out of this sector into lower spin subspaces, resulting in the decay of $\langle S^2 \rangle$, which imposes fundamental limits on spin squeezing. 
% This is clearly shown in Fig.~\ref{fig:Bloch_rotor_Lx_8_Ly_8}(a).
% The uncontrolled system (yellow data) reveals a characteristic decay of $\langle S^2\rangle$ along with oscillations. Strikingly, the controlled system~(blue data) shows dramatically suppressed decay of $\langle S^2\rangle$, demonstrating the field's
% active suppression of detrimental subspace transitions.
% This is further evidenced by the continuous squeezing growth under control~[Fig.~\ref{fig:Bloch_rotor_Lx_8_Ly_8}(b)] and the remarkable agreement between full RSW theory~(incorporating both collective spin dynamics and excitations) and the pure collective spin model, confirming the suppression of inter-subspace coupling. 

% \begin{figure}[h]
% {\includegraphics[width=\linewidth]{figure/Bloch_rotor_Lx_8_Ly_8_part_1_axis.png}
% \\
% \includegraphics[width=\linewidth]{figure/Bloch_rotor_Lx_8_Ly_8_part_2.pdf}}
% \caption{\label{fig:epsart}Optimized spin squeezing result for system size $8\times8$ and $T=1.5J^{-1}$. (a)-(b)Visualization of the spin squeezing dynamics in (c) on the Bloch spheres for the uncontrolled case and optimized case, respectively. We look down at the $y-z$ plane from the positive $x$-axis.(c)Time evolutions of the SSP using the OAT part in the RSW theory, and SSP using the full RSW theory (OAT+RSW). (d)The line shape of the optimized driving field $h(t)$ corresponding to (a)-(c). (e)An inset figure showing the time-evolution curves of the total angular momentum, and sharing the same legend as (c).}
% \label{fig:Bloch_rotor_Lx_8_Ly_8}
% \end{figure}

%The suppressed leakage into lower spin subspaces
The confinement of the system evolution predominantly within the maximal spin subspace
allows powerful visualization of the squeezing dynamics through the Husimi Q function~\cite{agarwal1981PRA}.
Defined as $Q(\theta,\phi)=|\langle\theta,\phi|\psi(t)\rangle|^2$, the Q function's distortion vividly captures the quantum state's evolution.
The uncontrolled system exhibits characteristic OAT dynamics~(see panel e): initial squeezing develops along an axis gradually deviating from $z$~[see (ii)], followed by over-twisting that degrades spin squeezing~[see (iii)]---as evidenced by the decay of the mean spin $|\langle \bm S \rangle|$ to zero~(orange data in panel c).
The optimized control field fundamentally alters this trajectory through squeezing-rotation synchronization~(panel f): during initial squeezing, the field induces a steady rotation around the $x$-axis~(panel d), aligning the squeezing direction with the $z$-axis when the system approaches its natural squeezing maximum~[see (iv)]. This controlled alignment effectively mitigates the over-twisting that would otherwise dominate the subsequent dynamics---as evidenced by the suppressed decay of the mean spin $|\langle \bm S \rangle|$~(blue data in panel c).
Then the field transitions to oscillations~(panel d, gray-shaded region), that serve dual purposes---their zero time-average maintains the $z$-alignment while their dynamical action leads to sustained squeezing~[see (v)].
Similar mechanism for enhanced spin squeezing is also observed in all-to-all interacting systems when subjected to properly designed pulse sequences~\cite{duan2013PRA}. 

% ~(which represents the quantum state's projection onto coherent spin states, $|\langle\theta,\phi|\psi(t)\rangle|^2$) on Bloch spheres (panels c-d), as the system remains dominantly in the collective spin subspace. 

\begin{figure}[h]
\includegraphics[width=\linewidth]{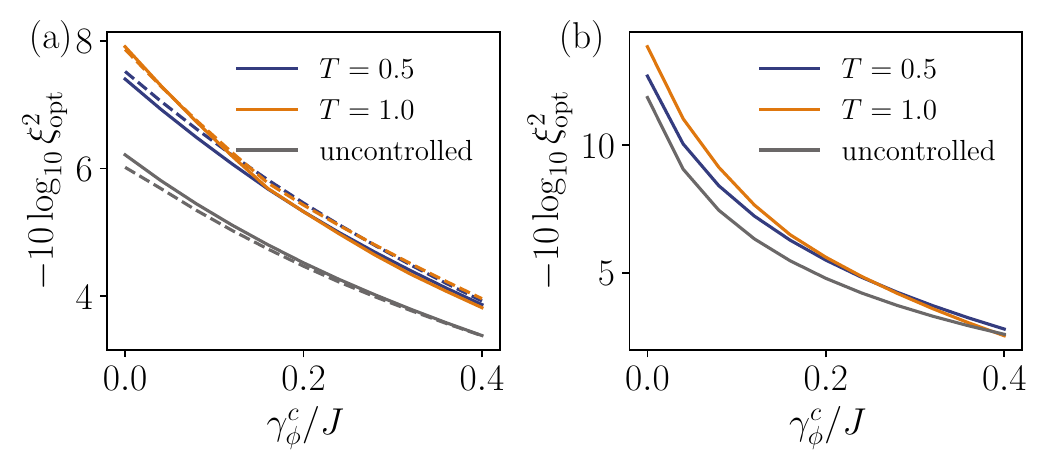}
\caption{The optimal squeezing parameter (in dB) as a function of the collective rate. (a) and (b) are for $4\times 4$ and $10\times 10$, respectively. Solid lines are the results from RSW theory. Gray curves denote the results for the uncontrolled case. Orange and blue curves denote the results for the optimized scheme with different total evolution time $T$. The dashed lines in (a) denote the results from ED. }
\label{fig:PBC_dephasing}
\end{figure}

% \begin{figure}[h]
% \includegraphics[width=\linewidth]{figure/collective_depahsing_PBC_ED_Lx_4_Ly_4_and_Lx_10_Ly_10_xlabel_dephasing.pdf}
% \caption{The effect of the collective dephasing on the optimizations in the PBC case. (a)Validation of the RSW theory in the case of noise by the ED method for system size $4\times 4$. The dots and the lines marked by 'x' represent the result obtained from ED and the RSW theory. The dashed-dotted lines represent the uncontrolled case under the dephasing noise. All the lines or markers correspond to $\gamma_\phi^c=0,0.08,0.16,0.24,0.32,0.4$ from above to below. (b)The RSW result for system size $10\times 10$.}
% \label{fig:PBC_dephasing}
% \end{figure}

% Having validated the control protocol for ideal conditions, we now probe its robustness against dephasing noise, which constitutes the dominant noise source in dipolar Rydberg atom arrays~\cite{Leseleuc2018AnalysisOI, wurtz2023aquilaqueras256qubitneutralatom}. 
\textit{Dephasing noise}---The enhanced squeezing in our scheme is achieved at the cost of a longer evolution time, which may reduce its effectiveness in realistic experimental settings where decoherence is unavoidable. In Rydberg atom arrays, a dominant decoherence mechanism is dephasing caused by fluctuations of the driving laser phase and atomic transition frequencies~\cite{Leseleuc2018AnalysisOI, wurtz2023aquilaqueras256qubitneutralatom}. 
To quantify the impact of decoherence on the control-enhanced squeezing dynamics, we focus on collective dephasing noise. This assumption enables large-scale simulations using the RSW theory. We emphasize, however, that the key physical insights are not tied to this specific noise model, and the effectiveness of the control protocol has also been verified in regimes with non-collective noise~(see Sec.~\ref{app:dephasing} of the Supplemental Material).
% Having established the control mechanism for ideal  systems, we next examine its resilience to noise. We consider collective phase noise---a dominant decoherence channel in many-body spin systems. 

Under collective dephasing noises, the system's dynamics follows the Lindblad master equation~\cite{Breuer2007open}:
\begin{equation}
    \partial_t\rho=-i[H,\rho] + \gamma_\phi^c \left(S_z \rho S_z-\frac{1}{2}\{S_z^2,\rho\}\right), \label{master_eq}
\end{equation}
where $\rho$ denotes the density matrix of the system and $\gamma_\phi^c$ denotes the dephasing rate. 
We generalize the RSW theory to open quantum systems by decomposing Eq.~\eqref{master_eq} into two components:
\begin{eqnarray}
    \partial_t [\rho]_\mathrm{ZM} && = -i\left[[H]_\mathrm{ZM},[\rho]_\mathrm{ZM}\right] \nonumber \\
    &&+\gamma_\phi^c\left(S_z [\rho]_\mathrm{ZM}S_z-\frac{1}{2}\{S_z^2,[\rho]_\mathrm{ZM}\}\right), \\
    \partial_t [\rho]_\mathrm{FM} && = -i[[H]_\mathrm{FM}, [\rho]_\mathrm{FM}].\label{master}
\end{eqnarray}
Here $[\rho]_\text{ZM}$ (zero-momentum sector) experiences the full dephasing noise, while $[\rho]_\text{FM}$ (finite-momentum sector) evolves unitarily. The explicit forms of $[H]_\text{ZM}$ (OAT Hamiltonian) and $[H]_\text{FM}$ (spin-wave Hamiltonian) can be found in Sec.~\ref{app:RSW} of the Supplemental Material.

Figure~\ref{fig:PBC_dephasing} 
%demonstrates the  robustness of our optimized squeezing protocol against collective dephasing noise, 
shows the optimal squeezing parameter as a function of the dephasing rate $\gamma_\phi^c$ for various total evolution time $T$.
%s ranging from $\gamma_\phi^c=0$ to $0.4J$. 
The ED~(dashed lines, obtained by Python library \texttt{Dynamiqs}~\cite{guilmin2025dynamiqs}) and RSW~(solid lines) results show excellent quantitative agreement for small systems (panel a), validating the RSW framework's accuracy and enabling its reliable extension to larger system sizes (panel b) where ED becomes computationally prohibitive.
Two key observations emerge from the results.
First, the controlled dynamics consistently achieve stronger squeezing than the uncontrolled case throughout the explored noise regime~($\gamma_\phi^c <0.4J$),demonstrating that the protocol retains a clear advantage even in the presence of substantial dephasing.
Second, for our protocol, a trade-off arises between squeezing generation and noise accumulation: longer evolution times 
$T$ produce stronger squeezing in the ideal limit~($\gamma_\phi^c=0$), but also increase the susceptibility to decoherence. As a result, optimal performance in noisy environments occurs at intermediate times where the coherent enhancement still dominates over decoherence effects. 
% The data reveals a crucial trade-off: while longer evolution times $T$ enable greater squeezing in ideal conditions~($\gamma_\phi^c=0$), they also offer more opportunity for noise to degrade performance. As a result, optimal performance in noisy environments occurs at intermediate times where the coherent enhancement still dominates over decoherence effects. Notably, while noise degrades absolute squeezing levels, 
% the controlled system consistently outperforms the uncontrolled case across all tested noise strengths.
% The advantage of the control protocol is expected to vanish when the noise strength exceeds the control's capability to enhance squeezing before decoherence dominates.

\begin{figure}[h]
\includegraphics[width=\linewidth]{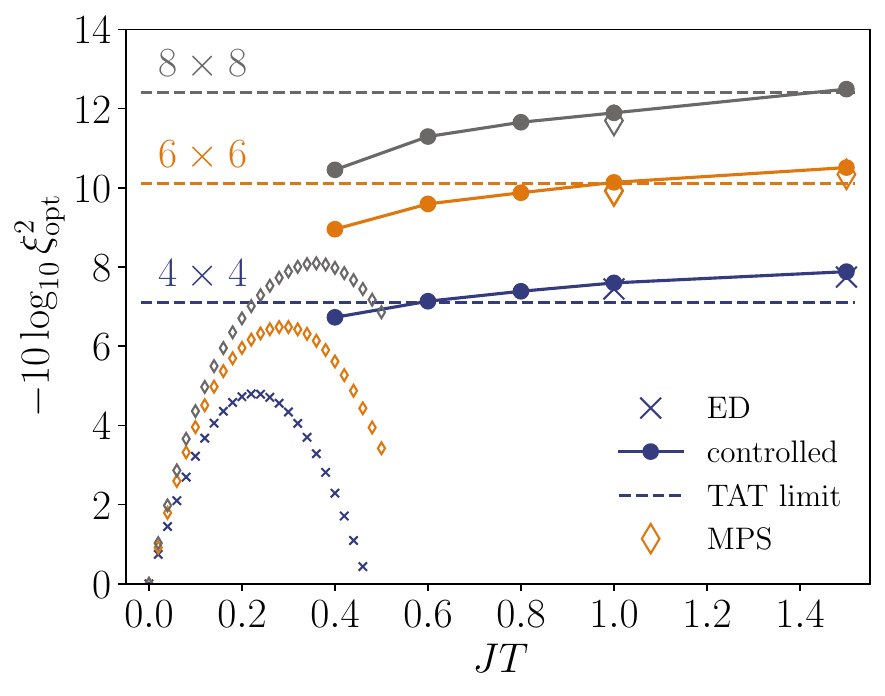}% Here is how to import EPS art
\caption{The optimal squeezing parameter for systems with open boundary conditions as a function of total evolution time (filled circles). The RSW theory generalized to OBCs is employed for the optimization procedure. In comparison, the time evolution of the squeezing parameter for the uncontrolled case are also shown, which are obtained using ED~(crosses) or MPS simulations~(diamonds) based on the time-dependent variational principle (TDVP, \texttt{ITensor}~\cite{Fishman_2022}), with a maximum bond dimension of $300$.
}
\label{fig:RSW_OBC}
\end{figure}

\textit{Open boundary condition}---Having established the efficacy of our protocol under periodic boundary conditions, we now turn to the more realistic setting of open boundary conditions. Although the rotor spin-wave (RSW) theory was originally formulated for translationally invariant systems \cite{roscilde2023rotor}, its underlying framework can be consistently generalized to OBCs through a suitable canonical transformation, as demonstrated in Sec.~\ref{app:RSW} of the Supplementary Material. 
% A detailed discussion of the generalization to OBCs and its verification is provided in Appendix~\ref{app:RSW} and Appendix~\ref{app:verification}.
For systems with OBCs, the accuracy of the RSW theory is slightly reduced due to the diminished interaction connectivity. Nevertheless, as demonstrated previously, the application of a control field effectively suppresses spin-wave excitations and preserves the validity of the rotor–spin separation. Consequently, the loss of accuracy observed in uncontrolled XX systems becomes negligible in the controlled setting. As shown in Fig.~\ref{fig:RSW_OBC}, the fields optimized within this protocol lead to a substantial enhancement of spin squeezing, demonstrating that the role of boundary conditions is primarily quantitative, while the underlying squeezing mechanism remains unchanged. The impact of noise is similar to that in the periodic-boundary case, as discussed in Sec.~\ref{app:dephasing} of the Supplemental Material.

% In the diagonalization of a quadratic bosonic Hamiltonian, one may, in certain cases, isolate a collective \textit{momentum} operator associated with a zero mode, corresponding to the zero-momentum sector in the RSW theory. However, the quadratic Hamiltonian alone does not yield the correct effective moment of inertia governing spin-squeezing dynamics (see Appendix~\ref{app:RSW}). Instead, this quantity must be obtained through the projection $\mathcal{P}_{J=N/2} H \mathcal{P}_{J=N/2}$, as prescribed in the RSW formalism.

% \cite{Colpa1986DiagonalizationOT, blaizot1986quantum, frerot2015area}

% As shown in Fig.~\ref{fig:RSW_OBC}, fields optimized using this protocol yield a substantial enhancement of spin squeezing, indicating that boundary conditions affect the results only quantitatively, while the underlying squeezing mechanism remains intact. Furthermore, in Appendix~\ref{app:dephasing}, we analyze the impact of dephasing noise, which constitutes the dominant noise source in dipolar Rydberg atom arrays\cite{Leseleuc2018AnalysisOI, wurtz2023aquilaqueras256qubitneutralatom}.

%\section{CONCLUSIONS}
\textit{Conclusion}---We have developed an optimal control strategy for generating enhanced spin squeezing in finite-range interacting systems. Focusing on a two-dimensional XX model with dipolar interactions, we demonstrated that optimizing a single collective transverse field is sufficient to substantially enhance squeezing, surpassing the conventional two-axis-twisting benchmark. By leveraging rotor–spin-wave theory for periodic boundary conditions, we were able to explore control protocols at large system sizes and identify the underlying physical mechanism: the optimized control dynamically suppresses interaction-induced mixing between spin sectors, thereby confining the dynamics predominantly within the maximal-spin subspace.
We further extended the theoretical framework to systems with open boundary conditions and dephasing noise, providing a scalable approach for analyzing realistic experimental settings. Our results show that the optimized protocol retains advantageous over uncontrolled dynamics across a broad noise regime, demonstrating its robustness against decoherence.

This work opens several promising directions for future research. In particular, the control principles identified here may be extended to other interaction geometries, higher-dimensional platforms, and more general many-body Hamiltonians. More broadly, combining optimal control with scalable many-body theories provides a powerful route for engineering entangled states in large quantum systems, paving the way toward practical quantum-enhanced sensing with programmable platforms.

\begin{acknowledgments}
We thank helpful discussions with Meng Khoon Tey, Zhenxing Hua, Yaofeng Chen, Peiyun Ge,  Xuanchen Zhang and Fan Yang. 
This work is supported by the National Key R\&D Program of China (Grant No.~2025YFA1411601) and NSFC (Grant No.~92576116).
Li You acknowledges support 
by NSFC (Grant No.~92265205).
\end{acknowledgments}



%\appendix
% \newpage
\clearpage
%\appendix

\section*{Supplemental Material}

\setcounter{figure}{0}
\renewcommand{\thefigure}{S\arabic{figure}}

% \large{\textbf{Supplemental material: Optimal Control of Spin Squeezing in 2D Finite-Range Interacting Systems}}\\

This supplemental material collects technical details that are not included in the main text. It is organized into three sections. Section~\ref{app:RSW} introduces the rotor spin-wave theory for systems with periodic and open boundary conditions, followed by its numerical validation in Sec.~\ref{app:verification}. Section~\ref{app:dephasing} investigates the robustness of the proposed protocol in the presence of individual dephasing noise.

\section{Rotor/spin-wave theory for quantum spin models with U(1) symmetry\label{app:RSW}}

Rotor/spin-wave theory provides a systematic and accurate approximation for describing the time evolution of quantum spin systems possessing U(1) symmetry~\cite{roscilde2023rotor}. Although the original paper only investigates the case of PBCs, their thoughts can consistently be generalized to that of OBCs by means of a general canonical transformation~\cite{Colpa1986DiagonalizationOT, blaizot1986quantum, frerot2015area}. In the first subsection, we will briefly summarize the original RSW theory following previous literature~\cite{roscilde2023rotor}, emphasizing the spin-$S$ XX model as an illustrative example. In the second subsection, we will demonstrate the treatment of OBCs. 

The Hamiltonian considered is
\begin{equation}
H = -\sum_{i<j} J_{ij}(S_i^x S_j^x + S_i^y S_j^y) - h(t)\sum_i S_i^x,
\end{equation}
where $S_i^\mu$ ($\mu=x,y,z$) denote the spin operators. For $S=1/2$, this Hamiltonian reduces to Eq.~(\ref{eq:XX_Ham}) in the main text.

The XX model supports a quantum easy-plane ferromagnet along the $xy$ plane with long-range order for $D\ge 2$~\cite{Frerot2017PRB}. Hence, we choose the $x$ axis as the quantization axis to perform the Holstein-Primakoff (HP) transformation~\cite{Holstein1940PR}:
\begin{eqnarray}
S_i^x&=&S-n_i,\nonumber \\
S_i^y&=&\frac{1}{2}\left(\sqrt{2S-n_i}a_i+a_i^\dagger\sqrt{2S-n_i}\right), \nonumber \\
S_i^z&=&\frac{1}{2i}\left(\sqrt{2S-n_i}a_i-a_i^\dagger\sqrt{2S-n_i}\right),
\end{eqnarray}
where $n_i=a_i^\dagger a_i$. Under this mapping, the initial state $|\mathrm{CSS}_x\rangle$ corresponds to the vacuum of HP bosons.

To linearize the transformation, we expand the square-root term in powers of $n_i/(2S)$ as follows:
\begin{equation}
\sqrt{2S-n_i} = \sqrt{2S}\left(1-\frac{n_i}{4S}\right) + \mathcal{O}(n_i^2),
\end{equation}
retaining only quadratic contributions while neglecting higher-order and constant terms. Consequently, we derive the exactly solvable quadratic Hamiltonian:
\begin{eqnarray}
    H_2 &=& -\sum_{i<j}J_{ij}S\bigg[-(n_i+n_j)+\frac{1}{2}(a_i a_j+a_i^\dagger a_j^\dagger) \nonumber \\
    & & + \frac{1}{2}(a_i^\dagger a_j + a_j^\dagger a_i)\bigg]- h(t)(NS-\sum_in_i).
    \label{eq:quadratic_Ham}
\end{eqnarray}
Up to now, our derivation applies to both PBCs and OBCs, and the difference in boundary conditions only affects the value of $J_{ij}$.

\subsection{Periodic boundary condition}
When the system is translational invariant, we can diagonalize the Hamiltonian $H_2$ by first performing the Fourier transformation
\begin{align}
a_i &= \frac{1}{\sqrt{N}}\sum_{\bm q} e^{i\bm q\cdot \bm r_i}a_{\bm q}, \\
J_{\bm q} &= \frac{1}{N}\sum_{ij}e^{i\bm q \cdot (\bm r_i - \bm r_j)}J_{ij}.
\end{align}
The Hamiltonian becomes
\begin{equation}
    H_2 = \frac{1}{2}\sum_{\bm q} \mqty(a_{\bm q}^\dagger \\ a_{-\bm q}) \mqty(A_{\bm q} & B_{\bm q} \\ B_{\bm q} & A_{\bm q}) \mqty(a_{\bm q} \\ a_{-\bm q}^\dagger) - \frac{1}{2}\sum_{\bm q}A_{\bm q},
\end{equation}
where
\begin{eqnarray}
    A_{\bm q} && = S[J_0-J_{\bm q} /2] +h(t),\\
    B_{\bm q} && = -J_{\bm q} S/2.
\end{eqnarray}
Subsequently, by applying the Bogoliubov transformation
\begin{equation}
b_{\bm q} = u_{\bm q} a_{\bm q} - v_{\bm q} a_{-\bm q}^\dagger,
\end{equation}
the quadratic Hamiltonian $H_2$ assumes the canonical form:
\begin{equation}
    H_2 = \sum_{\bm q}\epsilon_{\bm q}a_{\bm q}^\dagger a_{\bm q} + \frac{1}{2}\sum_{\bm q}(\epsilon_{\bm q} - A_{\bm q}),
    \label{eq:Harmonic}
\end{equation}
where $\epsilon_{\bm q}=\sqrt{A_{\bm q}^2-B_{\bm q}^2}$.

The dispersion relation has a zero mode $\epsilon_{\bm q}=0$ for $\bm q=0$, which leads to a singularity in the Bogoliubov transformation. 
The core concept of rotor/spin-wave theory~\cite{roscilde2023rotor} relies on the dominance of zero-momentum bosonic operators ($a_0$, $a_0^\dagger$) in the Hamiltonian, similar to Bose condensation but without approximating the zero-momentum mode as a classical field. Instead, it retains its full 
quantum nature by systematically isolating the 
$a_0$, $a_0^\dagger$-dependent terms in the Hamiltonian.

Specifically, the Hamiltonian $H_2$ is split into a zero-momentum component and a finite-momentum component, expressed as  
\begin{equation}  
    H_2 = [H]_{\text{ZM}}(a_0, a_0^\dagger) + [H]_{\text{FM}},  
\end{equation}  
where the zero-momentum term consists exclusively of the operators $a_0$ and $a_0^\dagger$, while the finite-momentum term comprises sums of products of bosonic operators, each involving at least one bosonic operator with finite momentum.  
% \begin{equation}
%     O=[O]_{\mathrm{ZM}}(b_0, b_0^\dagger) + [O]_{\mathrm{FM}}(b_{\bm q\neq 0}, b_{\bm q\neq 0}^\dagger).
% \end{equation}
% Here we use $b_{\bm q}$ instead of $a_{\bm q}$ because of the failure of the Bogoliubov transformation. 
Using the HP transformation for the $a_0$, $a_0^\dagger$ operators
\begin{eqnarray}
K_x&=&NS-n_0\nonumber, \\
K_y&=&\frac{1}{2}\left(\sqrt{2NS-n_0}a_0+\mathrm{H.c.}\right), \nonumber \\
K_z&=&\frac{1}{2i}\left(\sqrt{2NS-n_0}a_0-\mathrm{H.c.}\right),
\end{eqnarray}
the zero mode of $H$ can be easily cast into an OAT form 
\begin{align}
    [H]_\mathrm{ZM} &= \mathcal{P}_{J=NS}H\mathcal{P}_{J=NS} \nonumber \\
    &= \frac{K_z^2} {2I} -h(t)K_x+\mathrm{Const},
    \label{eq:H_ZM}
\end{align}
where 
\begin{equation}
    \frac{1}{2I}=\frac{J_0}{2(N-1)}
    \label{eq:inertia_PBC}
\end{equation}
gives the effective moment of inertia. Correspondingly, we have $[H]_\mathrm{FM} = H_2 - [H]_\mathrm{ZM}$.
% \begin{equation}
%     [H]_\mathrm{FM} = \frac{1}{2}\sum_{\bm q\neq 0} \mqty(b_{\bm q}^\dagger \\ b_{-\bm q}) \mqty(A_{\bm q} & B_{\bm q} \\ B_{\bm q} & A_{\bm q}) \mqty(b_{\bm q} \\ b_{-\bm q}^\dagger) - \frac{1}{2}\sum_{\bm q\neq 0}A_{\bm q},
% \end{equation}
% i.e., the $\bm q\neq0$ part of $H_2$.

Since in the second-order approximation Eq.~(\ref{eq:Harmonic}), different modes $\bm q$ do not couple with each other, the rotor Hamiltonian $ [H]_\mathrm{ZM}$ completely determines the equation of motion of $\bm q=0$ mode. The equation of motion for $\bm q\neq 0$ mode can also be easily obtained from Eq.~(\ref{eq:Harmonic})~\cite{roscilde2023rotor}. 

Based on the approximation above, we can approximate many observables. For example, the $x$ exponent of the total spin $\bm S$
\begin{equation}
    \langle \bm S_x \rangle = \langle K_x \rangle_R - \sum_{\bm q\neq 0}\langle a^\dagger_{\bm q} a_{\bm q}\rangle_{\mathrm{SW}},
    \label{eq:expecation_Sx}
\end{equation}
where $\langle K_x \rangle_R$ is given by the rotor Hamiltonian $[H]_\mathrm{ZM}$ and $N_\mathrm{FM} \equiv \sum_{\bm q\neq 0}\langle a^\dagger_{\bm q} a_{\bm q}\rangle_{\mathrm{SW}}$ is given by the $\bm q \neq 0$ mode in the quadratic Hamiltonian Eq.~(\ref{eq:Harmonic}). For the spin-squeezing parameter, there is also a simple formula
\begin{equation}
    \xi^2 \approx \frac{N \mathrm{min}_\theta \mathrm{Var}(K_\theta)_R}{(\langle K_x\rangle_R - N_\mathrm{FM})^2}. 
\end{equation}

\subsection{Open boundary condition}
% For open boundary systems, theoretical analysis proves that the rotor part may still exist in Eq.\ref{eq:quadratic_Ham} and can be decoupled from the spin waves.

For open boundary conditions, translational symmetry is absent, and momentum is no longer a good quantum number. 
As a result, the standard derivation based on Fourier modes does not apply, and it is \emph{a priori} unclear whether a well-defined collective rotor degree of freedom continues to exist.

The purpose of this subsection is to demonstrate that the rotor–spin-wave structure persists even in the absence of translational symmetry, albeit in a less transparent form.
We show that the quadratic spin-wave Hamiltonian~\eqref{eq:quadratic_Ham} under open boundary conditions still exhibits a single collective zero mode, which should be interpreted as the generator of global spin rotations.
Technically, this zero mode manifests itself through the non-diagonalizability of the quadratic Hamiltonian and requires a Jordan-block treatment.
Crucially, we emphasize that while the quadratic Hamiltonian correctly identifies the existence of the rotor mode, it does not by itself yield the correct effective moment of inertia.
The correct result should be obtained from the projection of the full Hamiltonian onto the maximal-spin sector as Eq.~\eqref{eq:H_ZM}, which is independent of boundary conditions.

To begin with, let us give the general procedure of diagonalizing a quadratic bosonic Hamiltonian and demonstrate its failure in the presence of a zero mode~\cite{blaizot1986quantum}. The Hamiltonian $H_2$ in Eq.~\eqref{eq:quadratic_Ham} can be written in a compact quadratic form
\begin{equation}
    H_2 = \sum_{i,j=1}^N a_i^\dagger A_{ij} a_j + \frac{1}{2}a_i^\dagger B_{ij}a_j^\dagger + \frac{1}{2}a_j B_{ji}^*a_i,
\end{equation}
where 
\begin{gather}
    A_{ij} =\delta_{ij}\left( S\sum_{k}J_{ik} + h(t)\right) - \frac{S}{2}J_{ij}, \label{Aij} \\ 
    B_{ij} = -\frac{S}{2}J_{ij}, \label{Bij}
\end{gather}
describe, respectively, the number-conserving and pairing terms arising from the Holstein–Primakoff expansion.
Introducing the Nambu spinor
\begin{gather}
\alpha = (a_1,\ldots,a_N,a_1^\dagger,\ldots,a_N^\dagger)^T,
    % a=(a_1,\dots,a_N)^T, \quad  \alpha = \mqty(a & a^\dagger)^T,
   % M =\mqty(A & B \\  B^* &  A^*)
\end{gather}
the Hamiltonian can be compactly expressed as
%we rewrite the Hamiltonian in the form
\begin{equation}
    H_2 = \frac{1}{2}\alpha^\dagger M \alpha - \frac{1}{2}\Tr A,
\end{equation}
with the Bogoliubov–de Gennes~(BdG) matrix
\begin{gather}
   M =\mqty(A & B \\  B^* &  A^*).
\end{gather}
 
To diagonalize $H_2$, one seeks a linear canonical transformation 
\begin{eqnarray}
    \alpha = T \beta,
\end{eqnarray}
where $\beta=(b_1,\ldots,b_N,b_1^\dagger,\ldots, b_N^\dagger)^T$, such that the bosonic commutation relations 
\begin{equation}
    [a_i,a_j]=[a_i^\dagger,a_j^\dagger]=0, \quad [a_i, a_j^\dagger]=\delta_{ij},
\end{equation}
are preserved. 
Writing the commutators in Nambu space as $$[\alpha,\alpha^\dagger]=\eta, \quad \eta=\mqty(\mathbb{I}_N & 0 \\ 0 & -\mathbb{I}_N)$$ with $\mathbb{I}_N$ being a $N$-dimensional identity matrix, this requirement implies
$T[\beta,\beta^\dagger]T^\dagger=\eta$. Since the transformed operators 
$\beta$ must satisfy the same bosonic algebra $[\beta,\beta^\dagger]=\eta$,
the transformation matrix $T$ must obey the paraunitary condition
\begin{eqnarray}
    T\eta T^\dagger=\eta. \label{T1}
\end{eqnarray}
In addition to preserving the bosonic commutation relations, a valid canonical transformation must also respect the Hermitian conjugation structure of the operators. In particular, the fundamental relation
$(a_i^\dagger)^\dagger = a_i$
must remain invariant under the transformation.
It can be written in matrix form as
\begin{equation}
\alpha^\dagger = (\gamma \alpha)^T,
\qquad
\gamma =
\begin{pmatrix}
0 & \mathbb{I}_N \\
\mathbb{I}_N & 0
\end{pmatrix}.\label{alpha} 
\end{equation}
This relation is a direct consequence of $(a_i^\dagger)^\dagger = a_i$ and encodes the conjugation structure in Nambu space.
The transformed operators $b_i$ are required to satisfy the same Hermitian conjugation relations as the original operators, namely
$(b_i^\dagger)^\dagger = b_i$.
In Nambu notation, this condition is equivalent to
\begin{equation}
\beta^\dagger = (\gamma \beta)^T. \label{beta}
\end{equation}
Combining the transformation $\alpha=T\beta$ with Eqs.~\eqref{alpha} and~\eqref{beta}, and using the identity $\gamma^2=I$, we obtain
\begin{eqnarray}
    T^* = \gamma T \gamma. \label{T2}
\end{eqnarray}
Equations~\eqref{T1} and~\eqref{T2} are two conditions on the transformation matrix $T$.

In terms of the transformed operators $\beta$,
% Introducing a new bosonic representation $\beta$ by
% \begin{equation}
%     \alpha=T\beta, \quad \beta=(b,b^\dagger)^T,
% \end{equation}
the Hamiltonian takes the form
\begin{equation}
    H_2=\frac{1}{2}\beta^\dagger \eta T^{-1} \eta M T \beta - \frac{1}{2}\Tr A .
\end{equation}
If the matrix $\eta M$ were diagonalizable, one could choose 
$T$ such that 
\begin{equation}
    T^{-1}\eta M T = \Omega,
\end{equation}
with $\Omega={\rm diag} (\omega_1,\ldots,\omega_N,-\omega_1,\ldots,-\omega_N)$. 
This pairing $\pm \omega_n$ in the spectrum is a direct consequence of particle–hole symmetry: if $(u,v)^T$ is an eigenvector with energy 
$E$, then $(v^*,u^*)^T$ is an eigenvector with energy $-E^*$
In the basis defined by $T$, the quadratic Hamiltonian takes the form
\begin{equation}
    H_2 = \frac{1}{2}\beta^\dagger \eta \Omega \beta - \frac{1}{2}\Tr A,
\end{equation}
% We can prove that the matrix $\Omega$ has the structure
% \begin{equation}
%     \Omega=\mqty(\omega & 0 \\ 0 & -\omega) \quad (\text{$\omega$: $N\times N$ diagonal}).
%     \label{eq:omega_structure}
% \end{equation}
which may be viewed as a collection of noninteracting quasiparticles:
\begin{equation}
    H=\sum_{n=1}^N \omega_n b_n^\dagger b_n  + \frac{1}{2}\sum_{n=1}^N \omega_n - \frac{1}{2}\Tr A.
\end{equation}

However, when a zero mode is present, the matrix $\eta M$ becomes non-diagonalizable and such a transformation does not exist. 
Actually, using the explicit forms of $A_{ij}$ and $B_{ij}$ from Eqs.~\eqref{Aij} and~\eqref{Bij} with $h(t)=0$,
one can verify:
$$\sum_j{(A_{ij}+B_{ij})}=0.$$
This suggests that the uniform vector
$(1,\ldots,1, 1,\dots,1)^T$ is an eigenvector of $\eta M$ with zero eigenvalue. 
Physically, this reflects that a uniform shift in the phase (or a uniform tilt in spin space) does not change the energy to quadratic order, which is precisely the signature of a Goldstone mode. The Goldstone mode appears as a zero-energy excitation, but due to particle-hole symmetry in the BdG formalism, the algebraic multiplicity of the zero eigenvalue is two. However, the uniform vector and its particle-hole conjugate (which is the same up to a phase here) are linearly dependent. Thus, the geometric multiplicity is only $1$, resulting in a Jordan block at zero energy. Consequently, the matrix 
$\eta M$ is non-diagonalizable and a complete Bogoliubov transformation does not exist. This signals the breakdown of the linear spin-wave description and necessitates a separate treatment of the zero mode, which is exactly the rotor/spin separation.

The procedure for separating the rotor and spin-wave components proceeds as follows. We begin with the full BdG matrix, written as  
\[
M = M_0 + h(t) \mathbb{I}_{2N},
\]  
where \(M_0\) corresponds to the Hamiltonian at zero transverse field \(h(t)=0\).
First, we examine the system in the absence of the external field. 
Let \(V^n\) denote the eigenvectors associated with {non-zero eigenvalues} \(\omega_n > 0\):
\[
\eta M_0 V^n = \omega_n V^n.
\]
Due to particle-hole symmetry of the bosonic BdG formalism, each $V_n$ has a partner  
\[
W^n = \gamma V^{n*},
\]  
which satisfies  
\[
\eta M_0 W^n = -\omega_n W^n.
\] 
In addition to these finite-frequency modes, the spectrum contains a zero-energy sector. Because the zero eigenvalue has algebraic multiplicity two but geometric multiplicity one, there exists a single (geometrically nondegenerate) zero-energy eigenvector $P$,
    \[
    \eta M_0 P = 0.
    \]
together with a generalized eigenvector 
$Q$ forming a Jordan chain,
    \[
    \eta M_0 Q = -\frac{i}{\mu} P,
    \]
where \(\mu > 0\) is a normalization constant. The factor \(-i/\mu\) is chosen so that \(P\) and \(Q\) can later be identified as a pair of canonically conjugate variables in the effective rotor description.
The action of $\eta M_0$
 in the zero-mode subspace can therefore be written as
\[
\eta M_0 \begin{pmatrix} P & Q \end{pmatrix} = \begin{pmatrix} P & Q \end{pmatrix}
\begin{pmatrix} 0 & -i/\mu \\ 0 & 0 \end{pmatrix},
\]
explicitly exhibiting the $2\times2$ Jordan block structure.

% We collect all eigenvectors into the transformation matrix
% \[
% T = \bigl( P,Q, V^1,\dots,V^{N-1},\; W^1,\dots,W^{N-1} \bigr).
% \]
% The first \(N\) columns contain the zero-mode eigenvector \(P\) and the \(N-1\) positive-frequency spin-wave modes \(V^n\); the last \(N\) columns contain the generalized zero-mode vector \(Q\) and the \(N-1\) negative-frequency partners \(W^n\).
To satisfy the conditions on $T$,~\eqref{T1} and~\eqref{T2}, the normalization relations can be chosen as ~\cite{blaizot1986quantum}
\begin{gather}
    V^{n\dagger }\eta V^m = -W^{n\dagger}\eta W^m = \delta_{mn}, \quad V^{n\dagger}\eta W^m=0, \label{eq: orthonormalization_1}\\
    W^n=\gamma V^{n*},\label{eq: orthonormalization_2}\\
    Q^\dagger M_0 Q=\frac{1}{\mu}, \quad Q^\dagger \eta P=i,\quad Q^\dagger \eta Q=0, \label{eq: orthonormalization_3}\\
    P = -\gamma P^*, \quad  Q=-\gamma Q^*. \label{eq: orthonormalization_4}
\end{gather}
Define two additional vectors $V^0$ and $W^0$ by
\begin{equation}
    V^0=\frac{1}{\sqrt{2}}(P+iQ),\quad W^0=-\frac{1}{\sqrt{2}}(P-iQ),
\end{equation}
then we can prove the relations
\begin{equation}
    V^{0\dagger}\eta V^0 = - W^{0^\dagger}\eta W^0 = 1, \quad V^{0^\dagger}\eta W^0 = 0.
\end{equation}
Hence, the relations~\eqref{eq: orthonormalization_1} and~\eqref{eq: orthonormalization_2} also hold when the vectors $V^0$ and $W^0$ are included in the set of vectors $V^n$ and $W^n$, and the matrix $T$ can be constructed as
\begin{equation}
    T = (V^0,\;V^1,\dots,V^{N-1},\;W^0,\;W^1,\dots,W^{N-1}).
\end{equation}
To present more intuitively, in the following text, we rearrange the column vectors of T in the form
\begin{equation}
    T = (V^0,\;W^0,\;V^1,\dots,V^{N-1},W^1,\dots,W^{N-1}).
\end{equation}
Correspondingly, the condition~\eqref{T1} now becomes 
\begin{equation}
    T \Tilde{\eta} T^\dagger = \eta, \quad \Tilde{\eta}=\mathrm{diag}(1,-1)\oplus \mqty(\mathbb{I}_{N-1} & 0 \\ 0 & -\mathbb{I}_{N-1}).
\end{equation}
The transformed matrix $T^{-1}\eta M_0 T$ thus takes a block-diagonal form consisting of the zero-mode Jordan block $\Omega_Z$ and the finite-frequency spectrum:
\begin{equation}
    T^{-1}\eta M_0 T = \Omega_Z \oplus \Omega_S,
    \label{eq:block_diagonal}
\end{equation}
where 
\begin{gather}
    \Omega_Z = \frac{1}{2\mu}\mqty(1 & 1 \\ -1 & -1), \\
    \Omega_S = \mathrm{diag}(\omega_1,\dots,\omega_{N-1},-\omega_1,\dots,-\omega_{N-1}).
\end{gather}

In terms of the corresponding quasiparticle operators, the Hamiltonian can be written as~\cite{blaizot1986quantum}
\begin{equation}
    H = \sum_{n>0}\omega_n b_n^\dagger b_n + \frac{\mathcal{P}^2}{2\mu} + \frac{1}{2}\sum_{n>0}\omega_n - \frac{1}{2}\Tr A,
\end{equation}
where $\mathcal{P}\equiv \alpha^\dagger \eta P$ plays the role of a momentum operator conjugate to the collective phase. 
For the Hamiltonian $H_2$ at $h(t)=0$, the zero mode is known to be proportional to $(1,1,\dots, 1)^T$.
After canonical normalization according to the above conditions, it can be chosen as
\[
P=-\frac{(i,i,\dots, i)^T}{\sqrt{2N}}.
\]
The corresponding momentum operator then reads
\begin{equation}
    \mathcal{P}=-\frac{i}{\sqrt{2N}}\sum_{i=1}^N (a_i - a_i^\dagger)=\sqrt{\frac{2}{N}}K^z,
\end{equation}
which yields the zero-mode contribution to the Hamiltonian,
\begin{equation}
    [H]_\mathrm{ZM}=\frac{1}{N\mu}(K^z)^2. 
\end{equation}
The same construction applies under periodic boundary conditions. Numerically, one finds $1/N\mu=J_0/2N$, in agreement with Eq.~(29) of Ref.~\cite{roscilde2023rotor}. 
However, as shown in that work, the correct moment of inertia is not given by this bare quadratic result, but by Eq.~\eqref{eq:inertia_PBC}. This confirms our main-text statement: the quadratic Hamiltonian alone does not yield the correct effective moment of inertia. Instead, it must be obtained by projecting onto the $J=NS$ sector,
$\mathcal{P}_{J=NS}H\mathcal{P}_{J=NS}$, as in Eq.~\eqref{eq:H_ZM}.

After isolating the rotor contribution, we proceed to determine the spin-wave sector.
For PBCs, translational invariance allows the spin-wave Hamiltonian to be diagonalized analytically in momentum space~(as shown in the previous section), yielding closed-form expressions for the Bogoliubov modes. In contrast, OBCs break translational symmetry, so that the spin-wave sector cannot be diagonalized analytically and must instead be treated by numerical diagonalization of the BdG matrix.
While the formal Bogoliubov transformation is well defined, its numerical implementation requires special care to ensure that the resulting quasiparticle operators satisfy the correct bosonic commutation relations. The numerical canonicalization of Bogoliubov eigenvectors is therefore a nontrivial task and has been discussed extensively in the literature
~\cite{Colpa1978DiagonalizationOT, Magnon_damping}. Among the canonical normalization conditions, Eqs.~\eqref{eq: orthonormalization_3} and~\eqref{eq: orthonormalization_4}, which involve only the zero-mode vectors, can be enforced straightforwardly by appropriate rescaling and phase fixing. In contrast, the conditions in Eqs.~\eqref{eq: orthonormalization_1} and~\eqref{eq: orthonormalization_2}, which constrain the mutual $\eta$-orthogonality of the spin-wave modes, capture the essential numerical subtleties and require special attention.
When all eigenvalues $\omega_n$
 are nondegenerate, standard numerical diagonalization routines already yield eigenvectors that satisfy the $\eta$-orthogonality conditions in Eq.~\eqref{eq: orthonormalization_1}. Consequently, no further processing is required in this case. The only nontrivial situation arises in degenerate subspaces, where numerical diagonalization returns arbitrary linear combinations of eigenvectors within the degenerate manifold. We therefore focus on a degenerate sector corresponding to a given eigenvalue $\omega$, spanned by the vectors
\[
\Tilde{S}=(\Tilde{V}^1,\dots,\Tilde{V}^M,\Tilde{W}^1,\cdots,\Tilde{W}^M).
\]
The goal is to construct, within this subspace, a new set of vectors that satisfy the canonical 
$\eta$-orthonormalization conditions. The canonicalization procedure within this degenerate sector proceeds as follows.
% Among the orthonormalization conditions, Eqs.~\eqref{eq: orthonormalization_3} and~\eqref{eq: orthonormalization_4} are comparatively straightforward to enforce.
% We therefore focus on Eqs.~\eqref{eq: orthonormalization_1} and~\eqref{eq: orthonormalization_2}, which already capture the essential numerical subtleties.
% When $\omega_n \neq \omega_m$, standard numerical diagonalization routines automatically yield eigenvectors satisfying Eq.~\eqref{eq: orthonormalization_1}.
% Consequently, the only nontrivial case arises in degenerate subspaces.
% We thus restrict our attention to a degenerate sector spanned by $\Tilde{S}=(\Tilde{V}^1,\dots,\Tilde{V}^M,\Tilde{W}^1,\cdots,\Tilde{W}^M)$.
% The canonicalization within this subspace proceeds as follows.
\begin{enumerate}
    \item Define $\Tilde{S}_+=(\Tilde{V}^1,\dots,\Tilde{V}^M)$ and compute the Gram matrix $G=\Tilde{S}_+^\dagger \eta \Tilde{S}_+$ with respect to the $\eta$-metric. The matrix $G$ is Hermitian and positive definite, as can be shown straightforwardly.
    \item Let $S_+=\Tilde{S}_+ G^{-1/2}$ and generate $S_-=\gamma S_+^*$.
    \item Form the new eigenvectors as $S=(S_+,S_-)=(V^1,\dots,V^M,W^1,\dots,V^M)$, which by construction satisfies Eqs.~\eqref{eq: orthonormalization_1} and~\eqref{eq: orthonormalization_2}.
\end{enumerate}

After determining the matrix decomposition of $\eta M_0$, we now turn to the case with a non-zero field $h(t)\neq0$. In the presence of $h(t)$, the block-diagonal structure of $T^{-1}\eta M T$ in~\eqref{eq:block_diagonal} no longer exist, but takes the form
\begin{equation}
    \Omega \equiv T^{-1}\eta M T = \mqty(\Omega_{ZZ} & \Omega_{ZS} \\ \Omega_{SZ} & \Omega_{SS} ),
\end{equation}
where $Z$ and $S$ represent the zero modes and the spin wave modes, respectively. To seperate the motion of the zero modes, the dynamics should be determined by $\Omega_{SS}$ in the spin-wave subspace. For example, expectation values associated with the spin-wave sector can be obtained by solving the equations of motion for the covariance matrix
\begin{equation}
    \expval{\alpha \alpha^\dagger} = \mqty( \mathbb{I}_N+D^* &  E \\ E^* & D),
    \label{eq:cov_alpha}
\end{equation}
where $D_{ij}=\expval*{a_i^\dagger a_j}$ and $E_{ij}=\expval{a_i a_j}$~\cite{frerot2015area}. Accordingly, we transform the covariance matrix to the $\beta$ representation by
\begin{equation}
    \expval{\beta \beta^\dagger} = T^{-1}\expval{\alpha\alpha^\dagger}(T^{-1})^\dagger=\mqty(C_{ZZ} & C_{ZS} \\ C_{SZ} & C_{SS}).
\end{equation}
Starting with the initial condition
\begin{align}
    \expval*{\beta \beta^\dagger}(0) & = T^{-1}\expval*{\alpha\alpha^\dagger}(0)(T^{-1})^\dagger \\
    & = T^{-1} \mqty(\mathbb{I}_N & 0_N \\ 0_N & 0_N) (T^{-1})^\dagger \\
    & \equiv \mqty( C_{ZZ}(0) & C_{ZS}(0) \\ C_{SZ}(0) & C_{SS}(0)),
\end{align}
we evolve the spin wave sector seperately by
\begin{equation}
    i\partial_t C_{SS}(t) = \Omega_{SS} C_{SS}(t) - C_{SS}(t) \Omega_{SS}^\dagger.
\end{equation}
At an arbitrary time $t$, we can recover the covariance structure in~\eqref{eq:cov_alpha} by
\begin{align}
    \expval{\alpha \alpha^\dagger}(t) &= T \mqty( C_{ZZ}(0) & C_{ZS}(0) \\ C_{SZ}(0) & C_{SS}(t)) T^\dagger \\
    & = \mqty(\mathbb{I}_{N} + D^*(t) & E(t) \\ E^*(t) & D(t) ),
\end{align}
from which the total particle number of the spin waves $\Tr D(t)$ is extracted.

There is another route that does not require transforming to the $\beta$ representation. Define
\begin{equation}
    T_S = (V^1,\dots,V^{N-1}, W^1,\dots,W^{N-1})
\end{equation}
and its left inverse
\begin{equation}
    T_S^{-1} = \eta_S T_S^\dagger \eta, \quad \eta_S = \mqty( \mathbb{I}_{N-1} & 0 \\ 0 & -\mathbb{I}_{N-1}).
\end{equation}
From Ref.~\cite{blaizot1986quantum}, we find that the operator $\Pi=T_S T_S^{-1}$ is a projector onto the spin-wave subspace, satisfying $\Pi^2 = \Pi$. Therefore, the generator 
\begin{equation}
    K = \Pi \eta M \Pi
\end{equation}
limits the dynamics within the spin-wave subspace. Similarly, we isolate the component of the initial state within the spin-wave subspace by 
\begin{equation}
    \expval{\alpha \alpha^\dagger}_S(0)=\Pi\expval{\alpha \alpha^\dagger}(0)\Pi^\dagger,
\end{equation}
and freeze the remaining component
\begin{equation}
    \expval{\alpha \alpha^\dagger}_\mathrm{remain}(0) = \expval{\alpha \alpha^\dagger}(0) - \expval{\alpha \alpha^\dagger}_S(0).
\end{equation}
The dynamics of $\expval{\alpha \alpha^\dagger}_S(t)$ is generated by
\begin{equation}
    i\partial_t \expval{\alpha \alpha^\dagger}_S(t) = K \expval{\alpha \alpha^\dagger}_S(t) - \expval{\alpha \alpha^\dagger}_S(t) K^\dagger.
\end{equation}
Supplemented with $\expval{\alpha \alpha^\dagger}_\mathrm{remain}(0)$, we obtain
\begin{equation}
    \expval{\alpha \alpha^\dagger}(t) = \expval{\alpha \alpha^\dagger}_S(t) + \expval{\alpha \alpha^\dagger}_\mathrm{remain}(0).
\end{equation}

Numerical verification of the method is demonstrated in the following section. %Fig.~\ref{fig:RSW_OBC} and Appendix.~\ref{app:verification}.

\section{Numerical Validation of RSW Theory\label{app:verification}}

\begin{figure}[h]
\includegraphics[width=\linewidth]{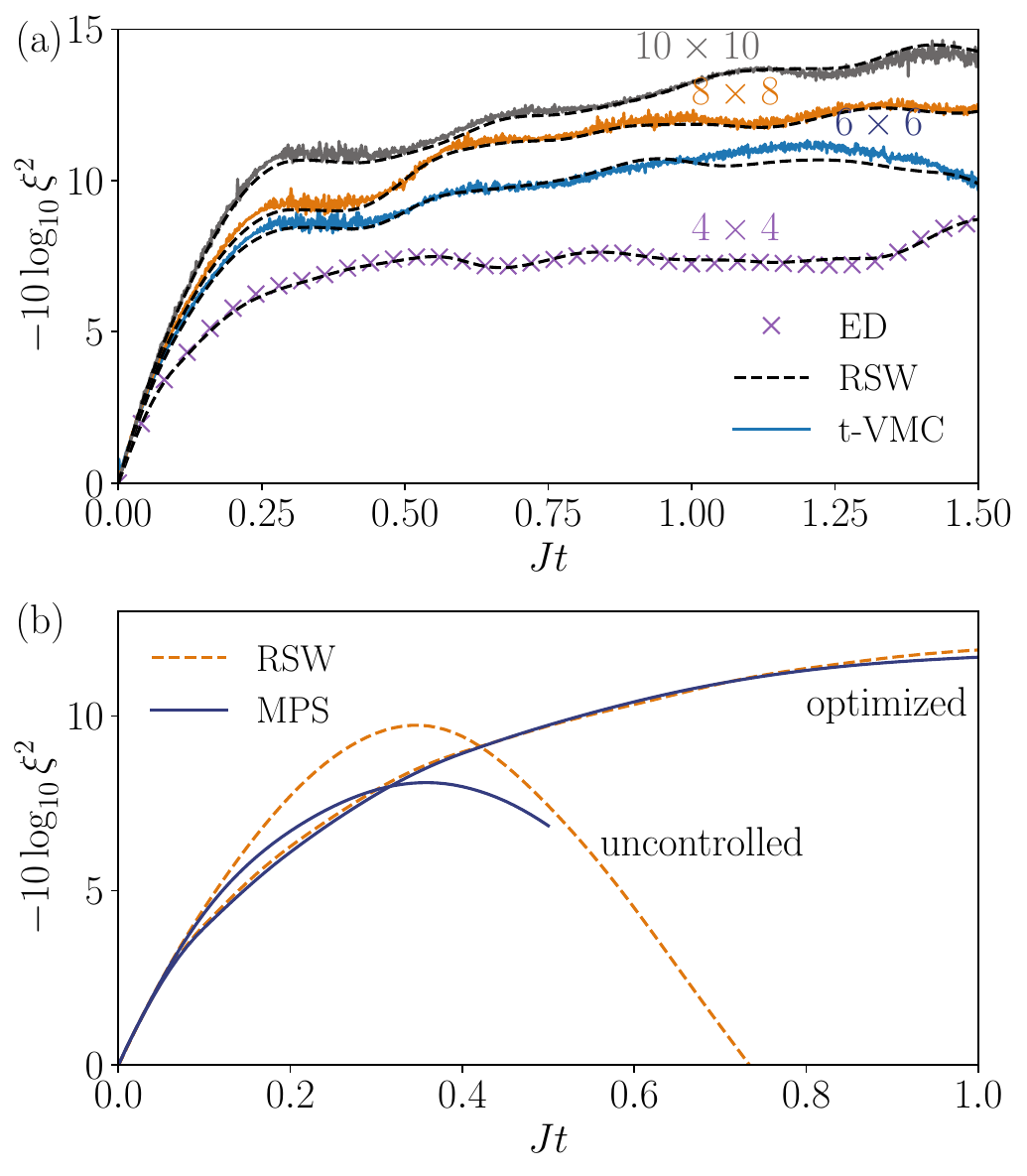}% Here is how to import EPS art
\caption{\label{fig:epsart} (a) The time evolution of the squeezing parameter for the control protocol for different system sizes with PBCs. The dashed curves are the RSW results. The solid curves are results from t-VMC. The ED results are denoted by crosses `x'. 
% Verification of the RSW theory by ED and the t-VMC method for different system sizes with PBCs. 
(b) The time evolution of the squeezing parameter for a $8\times 8$-size system with OBC. The dashed curves are RSW results. The solid curves are obtained by MPS (TDVP) method. 
% Verification of the RSW theory by MPS (TDVP) for $6\times 6$-size system with OBC. 
% The time evolution trajectories correspond to the $JT=1.0$ data point in Fig.~\ref{fig:RSW_OBC}.
}
\label{fig:compare_netket_and_rotor}
\end{figure}

For periodic boundary conditions, the validity of the RSW theory is established using two complementary numerical approaches. For small system sizes ($4 \times 4$), direct comparison with ED provides an exact benchmark. For larger systems, we employ t-VMC simulations with a translationally invariant restricted Boltzmann machine (RBM) ansatz~\cite{nomura2021helping}, implemented using the \texttt{NetKet} framework~\cite{netket3:2022}. As shown in Fig.~\ref{fig:compare_netket_and_rotor}(a), the excellent quantitative agreement between the numerical results (colored solid lines) and the RSW predictions (black dashed lines) confirms the validity of the theory across all relevant system sizes.

For open boundary conditions, the generalized RSW theory developed in Sec.~\ref{app:RSW} is benchmarked against MPS simulations, using the same methodology as in Fig.~\ref{fig:RSW_OBC}; the corresponding comparison is presented in Fig.~\ref{fig:compare_netket_and_rotor}(b). As discussed in the main text, the RSW theory generally exhibits reduced accuracy under OBCs. However, the degradation in accuracy observed in uncontrolled XX systems is significantly mitigated under the proposed control scheme. Consequently, the generalized RSW theory remains highly effective for simulating optimized spin-squeezing dynamics in large systems.\\

\section{Robustness to individual dephasing noise\label{app:dephasing}}
In this section, 
We examine the robustness of our protocol under individual dephasing noise by numerically solving the Lindblad master equation governing the system dynamics:
\begin{equation}
\partial_t\rho = -i[H,\rho] + \gamma_\phi \sum_i \left( S_i^z \rho S_i^z - \frac{1}{4}\rho \right),
\end{equation}
% where $\gamma_\phi$ denotes the dephasing strength. The simulations employ the quantum-jump approach through the \texttt{mcsolve} function in the \texttt{QuTip} quantum toolbox~\cite{johansson2012qutip,johansson2013qutip,lambert2024qutip5quantumtoolbox}.
The simulations employ the Monte Carlo wave function method implemented by ED~\cite{Molmer:93}.

\begin{figure}[h]
{
\includegraphics[width=\linewidth]{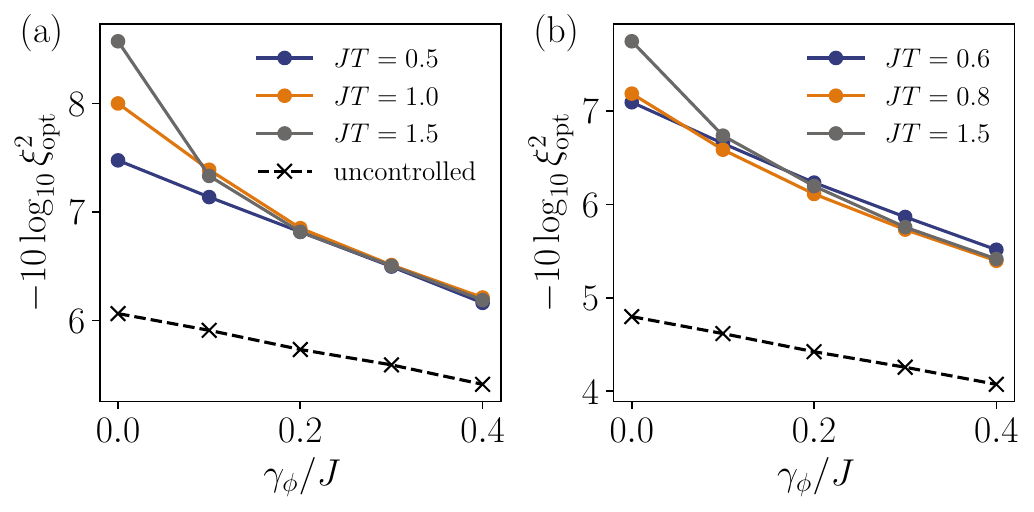}
}
\caption{The optimal squeezing parameter as a function of site-dephasing noise rate for the (a) PBC case and (b) OBC case, with $L_x=L_y=4$. 
Solid curves denote the results for the control protocol with various total evolution times.
The dashed curves denote the results for the uncontrolled case.
% Here we compare the optimization result for different $T$, to the uncontrolled case.
}
\label{fig:depahsing_OBC_Lx_4_Ly_4}
\end{figure}

As shown in Fig.~\ref{fig:depahsing_OBC_Lx_4_Ly_4}, the optimal squeezing parameter (in dB) demonstrates remarkable resilience against dephasing noise under both periodic~(see panel a) and open~(see panel b) boundary conditions. Notably, our protocol maintains superior performance compared to the uncontrolled case even at substantial dephasing strengths ($\gamma_\phi = 0.4$), although the advantage of longer evolution times $T$ becomes less pronounced with increasing noise. These findings exhibit similar characteristics to the collective dephasing case discussed in the main text, confirming the inherent noise tolerance of our approach across different decoherence channels.

\nocite{*}

\end{document}